\documentclass[12pt,draftcls,onecolumn]{IEEEtran}
\IEEEoverridecommandlockouts
\usepackage{cite}
\usepackage{amsmath,amssymb,amsfonts}
\usepackage{algorithmic}
\usepackage{graphicx}
\usepackage{textcomp}
\usepackage{xcolor}
\usepackage{amsthm}

\def\BibTeX{{\rm B\kern-.05em{\sc i\kern-.025em b}\kern-.08em
    T\kern-.1667em\lower.7ex\hbox{E}\kern-.125emX}}
\begin{document}

\title{On Existence of Girth-8 QC-LDPC Code with Large Column Weight: Combining Mirror-sequence with Classification Modulo Ten*\\
%{\footnotesize \textsuperscript{*}Note: Sub-titles are not captured for https://ieeexplore.ieee.org  and
%should not be used}
\thanks{This work was supported by the National Natural Science Foundation of
China (NSFC) under Grant 62571429.}
}

\author{%
   \IEEEauthorblockN{Guohua Zhang\IEEEauthorrefmark{1},
                     Xiangya Liu\IEEEauthorrefmark{1},
                     Jianhua Zhang\IEEEauthorrefmark{2},
                     and Yi Fang\IEEEauthorrefmark{3}}\\
   \IEEEauthorblockA{\IEEEauthorrefmark{1}%
                    School of Communications and Information Engineering, Xi'an University of Posts and Telecommunications, Xi'an, China,
                    zhangghcast@163.com}\\
   \IEEEauthorblockA{\IEEEauthorrefmark{2}%
                     China Academy of Space Technology (Xi'an), Xi'an, China,
                     zhangjhcast504@163.com}\\
   \IEEEauthorblockA{\IEEEauthorrefmark{3}%
                     School of Information Engineering, Guangdong University of Technology, Guangzhou, China,
                     fangyi@gdut.edu.cn}
 }
\maketitle

\begin{abstract}
Quasi-cyclic (QC) LDPC codes with large girths play a crucial role in several research and application fields, including channel coding, compressed sensing and distributed storage systems. A major challenge in respect of the code construction is how to obtain such codes with the shortest possible length (or equivalently, the smallest possible circulant size) using algebraic methods instead of search methods. The greatest-common-divisor (GCD) framework we previously proposed has algebraically constructed QC-LDPC codes with column weights of 5 and 6, very short lengths, and a girth of 8. By introducing the concept of a mirror sequence and adopting a new row-regrouping scheme, QC-LDPC codes with column weights of 7 and 8, very short lengths, and a girth of 8 are proposed for arbitrary row weights in this article via an algebraic manner under the GCD framework. Thanks to these novel algebraic methods, the lower bounds (for column weights 7 and 8) on consecutive circulant sizes are both improved by asymptotically about $20\%$, compared with the existing benchmarks. Furthermore, these new constructions can also offer circulant sizes asymptotically about $25\%$ smaller than the novel bounds.
\end{abstract}

\begin{IEEEkeywords}
circulant, exponent matrix, girth, quasi-cyclic (QC), low-density parity-check (LDPC).
\end{IEEEkeywords}

\section{Introduction}
In the field of channel coding, it is usually desired for a quasi-cyclic (QC) low-density parity-check (LDPC) code \cite{G25,JF06,ZZ14} to have the shortest possible length for a given girth. For the domain of compressed sensing \cite{DSV12,LXF17} or distributed storage systems \cite{Su18}, the parity-check matrix (PCM) of a QC-LDPC code with a large girth is also very popular, since it is a promising candidate for directly serving as a measurement matrix \cite{LXF17} or directly generating a fractional repetition code \cite{Su18}. For small column weights, such as 3 or 4, short QC-LDPC codes with girths larger than 6 have been constructed by certain well-designed search methods \cite{TBS17,KDMZ20} and algebraic methods \cite{ZSLF24}. Regarding medium column weights, such as 5 and 6, the existing search methods have shown signs of being inadequate, especially when the row weight is relatively large. By comparison, certain algebraic methods \cite{ZSNF24,ZSLF24} (such as the schemes based on greatest-common-divisor (GCD) framework \cite{ZSW13,ZZ14}) are still fairly competent for these column weights for arbitrary row weights. As the column weight continues to increase, the codes constructed by existing algebraic methods \cite{ZFL19,MGR22} become unsatisfactory because the resulting length is somewhat too long.

It should be noted that although QC-LDPC codes with large columns tend to have bad decoding thresholds, with the help of general techniques (such as masking and column splitting), they can be transformed into the counterparts with smaller column weights without reducing the girth. The all-zero blocks generated in the PCM after such operations are beneficial for improving the upper bound of the minimum distance, and therefore have significant potential to greatly enhance the performance in the error floor region. In addition, the codes with larger column weights are also good choices as components for constructing well-performing dual-diagonal QC-LDPC codes \cite{ZLSF24}.

By adopting the concept of a mirror sequence and a new row-regrouping scheme (modulo 10, instead of modulo 6 \cite{ZSNF24,ZSLF24}), QC-LDPC codes with column weights of 7 and 8 are algebraically proposed with very short lengths and a girth of 8 for arbitrary row weights in this article. The length of a QC-LDPC code considered here equals the row weight multiplied by the circulant size. Compared with the state-of-the-art works, the novel methods offer significantly enhanced lower bounds (about $20\%$ improvement) on consecutive circulant sizes. Moreover, the algebraic constructions we proposed can also provide noticeably smaller circulant sizes (about $25\%$ improvement) below these lower bounds for arbitrary row weights.

\section{Explicit Constructions for $J=7$}
Through extensive random search and pattern induction, we have identified a total of ten construction methods according to different values of $mod(L,10$). The principle we used to seek these methods is to make the maximum element of the mirror sequence (see explanation below) that defines the exponent matrix as small as possible. By further sorting out the relationships between these methods, they can be divided into a couple of basic constructions and many types of derived ones, with the latter being obtainable from the former. This section analyzes both the lower bound on consecutive circulant sizes and smaller circulant sizes below this bound for each of these methods. Finally, two general theorems on the circulant sizes are summarized for $(7,L)$-regular QC-LDPC codes without 4-cycles and 6-cycles.

\subsection{Two Basic Constructions and Their Derived Ones}

\subsubsection{First Group of Constructions}
In this part, we propose a basic construction, followed by a series of derived ones.

\emph{Basic Construction (J7-135)}: If $mod(L,10)\in\{1,3,5\}$, set $\textbf{E}=\textbf{M}^T\cdot[0,1,\cdots,L-1]$, where $\textbf{M}=[0,4,L,L+8,3L+4,3L+12,4L+8]$.

Because $4+(4L+8)=L+(3L+12)=(L+8)+(3L+4)$, the $\textbf{M}$ sequence involved in defining $\textbf{E}$ is termed as a \emph{mirror sequence}. From the above basic construction, the following construction can be readily obtained.

\emph{Derived Construction (J7-024)}: If $mod(L,10)\in\{0,2,4\}$, set
$\textbf{E}=[0,4,L+1,L+9,3L+7,3L+15,4L+12]^T\cdot[0,1,\cdots,L-1]$.

\emph{Proof}: First, by setting $L=L'+1$ in Basic Construction (J7-135) where $mod(L',10)\in\{0,2,4\}$, we have $\textbf{E}=[0,4,L'+1,L'+9,3L'+7,3L'+15,4L'+12]^T\cdot[0,1,\cdots,L'-1,L']$. Then, by deleting the last column and replacing $L'$ with $L$, the proof is completed. \qed

Similarly, the above construction in turn can be employed to generate a construction for the case of $mod(L,10)=9$.

\emph{Derived Construction (J7-9)}: If $mod(L,10)=9$, set
$\textbf{E}=[0,4,L+2,L+10,3L+10,3L+18,4L+16]^T\cdot[0,1,\cdots,L-1]$.

Once more, this construction can produce a construction for the scenario of $mod(L,10)=8$.

\emph{Derived Construction (J7-8)}: If $mod(L,10)=8$, set
$\textbf{E}=[0,4,L+3,L+11,3L+13,3L+21,4L+20]^T\cdot[0,1,\cdots,L-1]$.

\subsubsection{Second Group of Constructions} In this part, another basic construction is proposed, and then a derived one is presented.

\emph{Basic Construction (J7-7)}: If $mod(L,10)=7$, set $\textbf{E}=\textbf{M}^T\cdot[0,1,\cdots,L-1]$, where $\textbf{M}=[0,8,L,L+16,3L+8,3L+24,4L+16]$.

The basic construction can be utilized to generate a construction for the case of $mod(L,10)=6$.

\emph{Derived Construction (J7-6)}: If $mod(L,10)=6$, set $\textbf{E}=[0,8,L+1,L+17,3L+11,3L+27,4L+20]^T\cdot[0,1,\cdots,L-1]$.

\subsection{Lower Bound on Consecutive Circulant Sizes}
The characteristic of consecutive circulant sizes (CCS), refers to the concept that, with the exponent matrix remaining unchanged, the circulant size can change continuously (with a step of one) above a certain threshold. This attribute is crucial for studying the existence of QC-LDPC codes and for applications that require flexible lengths.

For the first basic construction, we have the following lemma regarding the lower bound on CCS.

\begin{table}[t]
\scriptsize
   \caption{Reason for the absence of 4-cycles and 6-cycles associated with every triple rows within the exponent matrix of Lemma 1. (the notation ``L odd-z'' means that L is an odd number satisfying $mod(L,10)\neq z$)}
   \label{t1}
   \centering
\begin{tabular}{l|l|l|l|l}
\hline
$\#$ & triple & reduced form & GCD indicator & reason\\\hline
1&[0,4,L]&-&L&L odd\\
2&[0,4,L+8]&-&L+8&L odd\\
3&[0,4,3L+4]&-&3L+4&L odd\\
4&[0,4,3L+12]&-&3L+12&L odd\\
5&[0,4,4L+8]&-&L+2&any L\\
6&[0,L,L+8]&[0,8,L+8]&L+8&L odd\\
7&[0,L,3L+4]&-&3L+4&L odd\\
8&[0,L,3L+12]&-&$\geq L+4$&L odd\\
9&[0,L,4L+8]&-&4L+8&L odd\\
10&[0,L+8,3L+4]&-&3L+4&L odd-7\\
11&[0,L+8,3L+12]&-&$\geq L+4$&L odd\\
12&[0,L+8,4L+8]&-&$\geq (4L+8)/3$&L odd\\
13&[0,3L+4,3L+12]&[0,8,3L+12]&3L+12&L odd\\
14&[0,3L+4,4L+8]&[0,L+4,4L+8]&4L+8&L odd\\
15&[0,3L+12,4L+8]&[0,L-4,4L+8]&$\geq (4L+8)/3$&L odd\\
16&[4,L,L+8]&[0,8,L+4]&L+4&L odd\\
17&[4,L,3L+4]&[0,L-4,3L]&$\geq L$&L odd\\
18&[4,L,3L+12]&[0,L-4,3L+8]&3L+8&L odd-9\\
19&[4,L,4L+8]&[0,L-4,4L+4]&4L+4&L odd-9\\
20&[4,L+8,3L+4]&[0,L+4,3L]&$\geq L$&L odd\\
21&[4,L+8,3L+12]&[0,L+4,3L+8]&3L+8&L odd\\
22&[4,L+8,4L+8]&[0,L+4,4L+4]&$\geq (4L+4)/3$&L odd\\
23&[4,3L+4,3L+12]&[0,8,3L+8]&3L+8&L odd\\
24&[4,3L+4,4L+8]&[0,L+4,4L+4]&same as \#22 &-\\	
25&[4,3L+12,4L+8]&[0,L-4,4L+4]&same as \#19&-\\	
26&[L,L+8,3L+4]&[0,8,2L+4]&L+2&L odd\\
27&[L,L+8,3L+12]&[0,8,2L+12]&L+6&L odd\\
28&[L,L+8,4L+8]&[0,8,3L+8]&same as \#23&-\\	
29&[L,3L+4,3L+12]&[0,8,2L+12]&same as \#27&-\\	
30&[L,3L+4,4L+8]&[0,L+4,3L+8]&same as \#21&-\\	
31&[L,3L+12,4L+8]&[0,L-4,3L+8]&same as \#18&-\\	
32&[L+8,3L+4,3L+12]&[0,8,2L+4]&same as \#26&-\\	
33&[L+8,3L+4,4L+8]&[0,L+4,3L]&same as \#20&-\\	
34&[L+8,3L+12,4L+8]&[0,L-4,3L]&same as \#17&-\\	
35&[3L+4,3L+12,4L+8]&[0,8,L+4]&same as \#16&-\\\hline
\end{tabular}
\end{table}

\emph{Lemma 1}: Let $mod(L,10)\in\{1,3,5\}$. For each circulant size $P\geq(L-1)(4L+8)+1$, the Tanner graph associated with $\textbf{E}$ in Basic Construction (J7-135) has a girth of eight.

%\emph{Proof}: See the full version of this article. \qed

\emph{Proof}: There are $C(7,3)=35$ triples. Firstly, we analyze four representative cases: case (4), case (5), case (10) and case (18).

The triple for case (4) is $[0,4,3L+12]$. Such a triple determines a $3\times L$ exponent matrix $\textbf{E}_{sub}$ which is composed of three rows selected from $\textbf{E}$. The GCD indicator of this triple is $(3L+12)/gcd(3L+12,4)$, which equals $(3L+12)/gcd(3L,4)=3L+12$ because $L$ is odd. This shows that the GCD constraint (GCD indicator being at least $L$) is satisfied. As a result, there are no 4-cycles and 6-cycles in the Tanner graph associated with $\textbf{E}_{sub}$.

The triple for case (5) is $[0,4,4L+8]$, and its GCD indicator is $(4L+8)/gcd(4L+8,4)=L+2$ for any value of $L$.

For case (10), the triple is $[0,L+8,3L+4]$. Since $L$ is odd, 2 is not a factor of $L+8$; similarly, 5 is not a factor of $L+8$ due to $mod(L,10)\notin\{2,7\}$. Therefore, the GCD indicator equals $(3L+4)/gcd(3L+14,L+8)=(3L+4)/gcd(-20,L+8)=3L+4$.

As for case (18), the triple is $[4,L,3L+12]$, which is equivalent to $[0,L-4,3L+8]$. Since $L$ is odd, 2 is not a factor of $L-4$; similarly, 5 is not a factor of $L-4$ due to $mod(L,10)\notin\{4,9\}$. Therefore, the GCD indicator equals $(3L+8)/gcd(3L+8,L-4)=(3L+8)/gcd(20,L-4)=3L+8$.

Secondly, every case whose number belongs to $\{1\sim4, 6\sim9, 11\sim17, 20\sim23, 26, 27\}$ is similar to case (4), and case (19) is similar to case (18). Finally, Each of the rest case (with number in the range $\{24,25,28\sim35\}$) is equivalent to a specific case which has been previously discussed. \qed

Regarding the second basic construction, we have the following lemma pertaining to the lower bound on CCS.

\begin{table}[t]
\scriptsize
   \caption{Reason for the absence of 4-cycles and 6-cycles associated with every triple rows within the exponent matrix of Lemma 2. (the notation ``L odd-z'' means that L is an odd number satisfying $mod(L,10)\neq z$)}
   \label{t2}
   \centering
\begin{tabular}{l|l|l|l|l}
\hline
$\#$ & triple & reduced form & GCD indicator & reason\\\hline
1&[0,8,L]&-&L&L odd\\
2&[0,8,L+16]&-&L+16&L odd\\
3&[0,8,3L+8]&-&3L+8&L odd\\
4&[0,8,3L+24]&-&3L+24&L odd\\
5&[0,8,4L+16]&-&L+4&L odd\\
6&[0,L,L+16]&[0,16,L+16]&L+16&L odd\\
7&[0,L,3L+8]&-&3L+8&L odd\\
8&[0,L,3L+24]&-&$\geq L+8$&L odd\\
9&[0,L,4L+16]&-&4L+16&L odd\\
10&[0,L+16,3L+8]&-&3L+8&L odd-9\\
11&[0,L+16,3L+24]&-&$\geq L+8$&L odd\\
12&[0,L+16,4L+16]&-&$\geq (4L+16)/3$&L odd\\
13&[0,3L+8,3L+24]&[0,16,3L+24]&3L+24&L odd\\
14&[0,3L+8,4L+16]&[0,L+8,4L+16]&4L+16&L odd\\
15&[0,3L+24,4L+16]&[0,L-8,4L+16]&$\geq (4L+16)/3$&L odd\\
16&[8,L,L+16]&[0,16,L+8]&L+8&L odd\\
17&[8,L,3L+8]&[0,L-8,3L]&$\geq L$&L odd\\
18&[8,L,3L+24]&[0,L-8,3L+16]&3L+16&L odd-3\\
19&[8,L,4L+16]&[0,L-8,4L+8]&4L+8&L odd-3\\
20&[8,L+16,3L+8]&[0,L+8,3L]&$\geq L$&L odd\\
21&[8,L+16,3L+24]&[0,L+8,3L+16]&3L+16&L odd\\
22&[8,L+16,4L+16]&[0,L+8,4L+8]&$\geq (4L+8)/3$&L odd\\
23&[8,3L+8,3L+24]&[0,16,3L+16]&3L+16&L odd\\
24&[8,3L+8,4L+16]&[0,L+8,4L+8]&same as \#22&-\\
25&[8,3L+24,4L+16]&[0,L-8,4L+8]&same as \#19&-\\
26&[L,L+16,3L+8]&[0,16,2L+8]&L+4&L odd\\
27&[L,L+16,3L+24]&[0,16,2L+24]&L+12&L odd\\
28&[L,L+16,4L+16]&[0,16,3L+16]&same as \#23&-\\
29&[L,3L+8,3L+24]&[0,16,2L+24]&same as \#27&-\\
30&[L,3L+8,4L+16]&[0,L+8,3L+16]&same as \#21&-\\
31&[L,3L+24,4L+16]&[0,L-8,3L+16]&same as \#18&-\\
32&[L+16,3L+8,3L+24]&[0,16,2L+8]&same as \#26&-\\
33&[L+16,3L+8,4L+16]&[0,L+8,3L]&same as \#20&-\\
34&[L+16,3L+24,4L+16]&[0,L-8,3L]&same as \#17&-\\
35&[3L+8,3L+24,4L+16]&[0,16,L+8]&same as \#16&-\\\hline
\end{tabular}
\end{table}

\emph{Lemma 2}: Let $mod(L,10)=7$. For each circulant size $P\geq(L-1)(4L+16)+1$, the Tanner graph associated with $\textbf{E}$ in Basic Construction (J7-7) has a girth of eight.

%\emph{Proof}: See the full version of this article. \qed

\emph{Proof}: There are $C(7,3)=35$ triples. Firstly, we analyze three representative cases: case (6), case (10) and case (18).

The triple for case (6) is $[0,L,L+16]$ which is equivalent to $[0,16,L+16]$. Because $L$ is odd, we have $gcd(L,16)=1$. The GCD indicator of this triple is $(L+16)/gcd(L+16,16)$, which equals $(L+16)/gcd(L,16)=L+16$.

For case (10), the triple is $[0,L+16,3L+8]$. Since $L$ is odd, 2 is not a factor of $L+16$; similarly, 5 is not a factor of $L+16$ due to $mod(L,10)\notin\{4,9\}$. Therefore, the GCD indicator equals $(3L+8)/gcd(3L+8,L+16)=(3L+8)/gcd(-40,L+16)=3L+8$.

As for case (18), the triple is $[8,L,3L+24]$, which is equivalent to $[0,L-8,3L+16]$. Since $L$ is odd, 2 is not a factor of $L-8$; similarly, 5 is not a factor of $L-8$ due to $mod(L,10)\notin\{3,8\}$. Therefore, the GCD indicator equals $(3L+16)/gcd(3L+16,L-8)=(3L+16)/gcd(40,L-8)=3L+16$.

Secondly, every case whose number belongs to $\{1\sim9, 11\sim17, 20\sim23, 26, 27\}$ is similar to case (6); moreover, case (19) is similar to case (18). Finally, Each of the rest case (with number in the range $\{24,25,28\sim35\}$) is equivalent to a specific case which has been previously discussed. \qed

\emph{Remark 1}: Assume that the mirror sequence $\textbf{M}$ is an integer sequence that satisfies the GCD constraint for the row weight $L$. Then $\textbf{E}=\textbf{M}^T\cdot[0,1,\cdots,L-1]$ corresponds to a QC-LDPC code without 4-cycles and 6-cycles for each circulant size $P\geq\textbf{M}(J-1)\cdot(L-1)+1$ \cite{ZSW13}. Obviously, $\textbf{M}$ naturally fulfill the GCD constraint for the row weight $L'=L-1$. Therefore, the exponent matrix $\textbf{E}'=\textbf{M}^T\cdot[0,1,\cdots,L'-1]$ (which is just the matrix obtained by deleting the last column of $\textbf{E}$) corresponds to a QC-LDPC code without 4-cycles and 6-cycle for each $P\geq\textbf{M}(J-1)\cdot(L'-1)+1$. By referring to Section II. A, it is readily seen that the values of $\textbf{M}(J-1)\cdot(L-1)+1$ for DC(J7-024), DC(J7-9), DC(J7-8) and DC(J7-6) are $(4L+12)(L-1)+1$, $(4L+16)(L-1)+1$, $(4L+20)(L-1)+1$ and $(4L+20)(L-1)+1$, respectively.

From Lemmas $1\sim2$ and Remark 1, a theorem follows immediately regarding the existence of $(7,L)$-regular QC-LDPC code without 4-cycles and 6-cycles.

\emph{Theorem 1}: When $L>7$, there exist $(7,L)$-regular QC-LDPC codes with a girth at least eight for each circulant size $P\geq P_{LB}=4(L-1)(L+5)+1$.

\subsection{Circulant Size Smaller Than Lower Bound}
The following two properties are useful for proving the theorem in this section.

P1\cite{ZSW13}: If a triple $[0,a_1,a_2]$ satisfies $0<a_1<a_2$ and $a_2/gcd(a_2,a_1)\geq L$, then the exponent matrix $[0,a_1,a_2]^T\cdot[0,1,\cdots,L-1]$ corresponds to a girth-8 QC-LDPC code for each circulant size $P\geq a_2(L-1)+1$.

P2\cite{ZSNF24}: Let $a_1$, $a_2$ and $b$ be three positive integers such that $b\geq L$, $gcd(a_1,b)=b$ and $gcd(a_2,b)=1$. Then the Tanner graph associated with the exponent matrix $[0,a_1,a_2]^T\cdot[0,1,\cdots,L-1]$ has no 6-cycles for each circulant size $P$ satisfying $gcd(P,b)=b$.

A triple has three common equivalent forms: (i)`S' operation: $(a_0,a_1,a_2)$ is equivalent to $(a_0-a_0,a_1-a_0,a_2-a_0)$ ; (ii) `R' operation: $(0,a_1,a_2)$ is equivalent to $(0,a_2-a_1,a_2)$; and (iii) `$/d$' operation: $(a_0*d,a_1*d,a_2*d)$ is equivalent to $(a_0,a_1,a_2)$, where $d$ and the circulant size are coprime.

\begin{table}[t]
\scriptsize
   \caption{Reason for the absence of 4-cycles and 6-cycles associated with every three rows within the exponent matrix of Lemma 3. }
   \label{t3}
   \centering
\begin{tabular}{l|l|l|l}
\hline
$\#$ & triple & equivalent form & reason\\\hline
1&[0,4,L]&-& P1\\
2&[0,4,L+8]&-& P1\\
3&[0,4,3L+4]&-& P1\\
4*&[0,4,3L+12]&-& need proof\\
5&[0,4,4L+8]&`/4'~[0,1,L+2]& P1\\
6&[0,L,L+8]&`R'~[0,8,L+8]& P1\\
7&[0,L,3L+4]&-& P1\\
8*&[0,L,3L+12]&-&need proof\\
9&[0,L,4L+8]&-& P2\\
10&[0,L+8,3L+4]&-&P1\\
11*&[0,L+8,3L+12]&-&need proof\\
12&[0,L+8,4L+8]&`R'~[0,3L,4L+8]& P2\\
13&[0,3L+4,3L+12]&-& P2\\
14&[0,3L+4,4L+8]&- & P2\\
15*&[0,3L+12,4L+8]&`R'~[0,L-4,4L+8]&need proof\\
16&[4,L,L+8]&`SR'~[0,8,L+4]&P1\\
17&[4,L,3L+4]&`S'~[0,L-4,3L]&P1\\
18*&[4,L,3L+12]&`S'~[0,L-4,3L+8]&need proof\\
19*&[4,L,4L+8]&`S'~[0,L-4,4L+4]&need proof\\
20&[4,L+8,3L+4]&`S'~[0,L+4,3L]&P1\\
21*&[4,L+8,3L+12]&`S'~[0,L+4,3L+8]&need proof\\
22&[4,L+8,4L+8]&`SR'~[0,3L,4L+4]& P2\\
23&[4,3L+4,3L+12]&`S'~[0,3L,3L+8]& P2\\
24&[4,3L+4,4L+8]&`SR'~[0,L+4,4L+4]&same as \#22\\	
25&[4,3L+12,4L+8]&`SR'~[0,L-4,4L+4]&same as \#19\\	
26&[L,L+8,3L+4]&`S/2'~[0,4,L+2]&P1\\
27&[L,L+8,3L+12]&`S/2'~[0,4,L+6]&P1\\
28&[L,L+8,4L+8]&`S'~[0,8,3L+8]&same as \#23\\	
29&[L,3L+4,3L+12]&`SR/2'~[0,4,L+6]&same as \#27\\	
30&[L,3L+4,4L+8]&`SR'~[0,L+4,3L+8]&same as \#21\\	
31&[L,3L+12,4L+8]&`SR'~[0,L-4,3L+8]&same as \#18\\	
32&[L+8,3L+4,3L+12]&`SR/2'~[0,4,L+2]&same as \#26\\	
33&[L+8,3L+4,4L+8]&`SR'~[0,L+4,3L]&same as \#20\\	
34&[L+8,3L+12,4L+8]&`SR'~[0,L-4,3L]&same as \#17\\	
35&[3L+4,3L+12,4L+8]&`S'~[0,8,L+4]&same as \#16\\\hline
\end{tabular}
\label{tab1}
\end{table}

\emph{Lemma 3}: If $mod(L,10)\in\{1,3\}$, the Tanner graph associated with exponent matrix of BC(J7-135) and the circulant size $P=L(3L+4)$ has a girth of eight.

\emph{Proof}: It is easy to prove the absence of 4-cycles, and hence the proof is omitted. Now we consider 6-cycles. There are $C(7,3)=35$ cases for 6-cycles, as listed in Table~\ref{tab1}. Among them, there are 18 cases which can be easily analyzed by the above two properties (P1 and P2), and there are 10 cases, each of which has a one-to-one equivalence relationship with another set of 10 cases. Therefore, there are only 7 cases left that need to be proven separately. In fact, we only prove three cases here: Case 4, Case 19 and Case 21, which cover the main skills used in the whole proof process. The rest four cases can be similarly proved and hence omitted here for brevity.

First, consider Case 4, the triple of $[0,4,3L+12]$. Suppose that 6-cycles exist. Then such cycles can be expressed as $(0-4j)+[4i-(3L+12)i]+[(3L+12)k-0]=nL(3L+4)$, which reduces to $(3L+4)(k-i)+4(2k-i-j)=nL(3L+4)$. So, $4(2k-i-j)=z(3L+4)$. Because $2*0-(L-1)-(L-2)\leq 2k-i-j\leq 2(L-1)-0-1$, we have $|2k-i-j|\leq 2L-3$. Therefore, $z\in\{0,\pm1,\pm2\}$. If $z=0$, then $k-i=nL$, which is impossible owing to $k\neq i$ and $|k-i|<L$. If $z=\pm1$, then $4(2k-i-j)=\pm(3L+4)$, which is impossible because $3L+4$ is odd and hence cannot be divided by 4. If  $z=\pm2$, then $4(2k-i-j)=\pm2(3L+4)$, which is impossible because $3L+4$ is odd and hence cannot be divided by 2. Consequently, we conclude that there are no 6-cycles for Case 4.

Next, consider Case 19, the triple of $[0,L-4,4L+4]$. Such cycles can be expressed by $[0-(L-4)j]+[(L-4)i-(4L+4)i]+[(4L+4)k-0]=nL(3L+4)$, which is $L(i-j)+4L(k-i)+4(k+j-2i)=nL(3L+4)$. It follows that $4(k+j-2i)=zL$. Since $|k+j-2i|\leq (2L-3)$ and $L$ is odd, we have $z\in\{0,\pm4\}$. (i) If $z=0$, then $k+j-2i=0$ and hence $(i-j)+4(k-i)=5(k-i)=n(3L+4)$. The following analysis is divided into three scenarios based on the value of $n$. $n=0$ leads to $k=i$, which is impossible; $n=\pm1$ leads to $k-i=\pm(3L+4)/5$, also impossible since $mod(L,10)\notin\{2,7\}$ and hence 5 cannot divides $\pm(3L+4)$; $n\geq 2$ cannot occur due to $|k-i|<L$. (ii) if $z=\pm4$, then $k+j-2i=\pm L$ and hence $(i-j)+4(k-i)\pm4=n(3L+4)$. By replacing $(i-j)$ with $k-i\mp L$, we obtain $5(k-i)=n(3L+4)\pm L\mp4$. As $|5(k-i)|\leq 5(L-1)$, it follows that $n\in\{0,\pm1\}$. Therefore, $5(k-i)\in\{\pm(L-4),\pm4L,\pm2(L-4)\}$. However, 5 cannot divides $(L-4)$ because $mod(L,10)\notin\{4,9\}$, and 5 cannot divides $4L$ because $mod(L,10)\notin\{0,5\}$. As a result, we conclude that there are no 6-cycles for Case 19.

Finally, consider Case 21, the triple of $(0,L+4,3L+8)$. It is easy to see that $gcd(3L-2,L(3L+4))=gcd(3L-2,6L)=gcd(3L-2,4)=1$. So, $(0,L+4,3L+8)$ is equivalent to $(3L-2)\cdot(0,L+4,3L+8)$ modulo $P$, which is $(0,6L-8,6L-16)$. It is equivalent to $(0,3L-4,3L-8)$ via '/2' operation, which is further turned into $(0,4,3L-8)$ via `R' operation. Therefore, there are no 6-cycles for Case 21. \qed

 %[0,4,L,L+8,3L+4,3L+12,4L+8]
%
%L(3L+4)

By a similar reasoning as in the proof of Lemma 3, the following four lemmas can be proved. The detailed proofs are omitted for the sake of brevity.

\emph{Lemma 4}: If $mod(L,10)=4$, the Tanner graph associated with the exponent matrix of DC(J7-024) and the circulant size $P=3L^2+10L+15$ has a girth of eight.

\emph{Lemma 5}: If $mod(L,10)=5$, the Tanner graph associated with the exponent matrix of BC(J7-135) and the circulant size $P=L(3L+6)$ has a girth of eight.

\emph{Lemma 6}: If $mod(L,10)=7$, the Tanner graph associated with the exponent matrix of BC(J7-7) and the circulant size $P=L(3L+8)$ has a girth of eight.

\emph{Lemma 7}: If $mod(L,10)=9$, the Tanner graph associated with the exponent matrix of DC(J7-9) and the circulant size $P=L(3L+10)$ has a girth of eight.

\emph{Remark 2}: In Lemma 3, choose the first $L-1$ columns of $\textbf{E}$. Then the obtained matrix corresponds to a $(7,L-1)$-regular QC-LDPC code with a girth of eight for the circulant size of $P=L(3L+4)$. By setting $L=L'+1$ where $mod(L',10)\in\{0,2\}$, it is easy to see that $(7,L')$-regular QC-LDPC code with a girth of eight can be constructed for $P=(L'+1)(3L'+7)$. Similarly, according to Lemma 6, $(7,L')$-regular QC-LDPC code with a girth of eight can be constructed for $P=(L'+1)(3L'+11)$, where $mod(L',10)=6$. Again, thanks to Lemma 7, $(7,L')$-regular QC-LDPC code with a girth of eight can be generated for $P=(L'+1)(3L'+13)$, where $mod(L',10)=8$.

From Lemmas $3\sim7$ and Remark 2, a theorem follows immediately regarding the existence of $(7,L)$-regular QC-LDPC codes without 4-cycles and 6-cycles. This theorem provides an upper bound on the minimum circulant size.

\emph{Theorem 2}: For each row weight $L>7$, the minimum circulant size $P_s$ for $(7,L)$-regular QC-LDPC codes with a girth at least eight satisfies $P_s\leq P_{UB}=(L+1)(3L+13)$.

According to Theorem 2, if one needs to find out the minimum circulant size for a $(7,L)$-regular QC-LDPC code without 4-cycles and 6-cycles, the search space can be narrowed down to the range between $6(L-1)+1$ and $P_{UB}$.

\section{Explicit Constructions for $J=8$}
Through extensive random search and pattern induction, we have also identified a total of ten construction methods for $J=8$ according to different values of $mod(L,10$). By further sorting out the relationships between these methods, they can be divided into a couple of basic constructions and many types of derived ones. This section analyzes both the lower bound on consecutive circulant sizes and smaller circulant sizes below this bound for each of these methods. Finally, two general theorems on the circulant sizes are summarized for $(8,L)$-regular QC-LDPC codes without 4-cycles and 6-cycles.

\subsection{Two Basic Constructions and Their Derived Ones}
\subsubsection{First Group of Constructions} In this part, we propose a basic construction, followed by a series of derived ones.

\emph{Basic construction (J8-135)}: If $mod(L,10)\in\{1,3,5\}$, then choose $\textbf{E}=\textbf{M}^T\cdot[0,1,\cdots,L-1]$, where $\textbf{M}=[0,4,L,L+8,3L+4,3L+12,4L+8,4L+12]$.

This construction leads to the following construction.

\emph{Derived construction (J8-024)}: If $mod(L,10)\in\{0,2,4\}$, then choose $\textbf{E}=[0,4,L+1,L+9,3L+7,3L+15,4L+12,4L+16]^T\cdot[0,1,\cdots,L-1]$.

Similarly, the above construction in turn can generate a construction for the case of $mod(L,10)=9$.

\emph{Derived construction (J8-9)}: If $mod(L,10)=9$, then choose $\textbf{E}=[0,4,L+2,L+10,3L+10,3L+18,4L+16,4L+20]^T\cdot[0,1,\cdots,L-1]$.

Once more, the above construction leads to a construction for the scenario of $mod(L,10)=8$.

\emph{Derived construction (J8-8)}: If $mod(L,10)=8$, then choose $\textbf{E}=[0,4,L+3,L+11,3L+13,3L+21,4L+20,4L+24]^T\cdot[0,1,\cdots,L-1]$.
\subsubsection{Second Group of Constructions} In this part, another basic construction is proposed, and then a derived one is presented.

\emph{Basic construction (J8-7)}: If $mod(L,10)=7$, then set $\textbf{E}=\textbf{M}^T\cdot[0,1,\cdots,L-1]$, where $\textbf{M}=[0,8,L,L+16,3L+8,3L+24,4L+16,4L+24]$.

The basic construction can generate a construction for the case of $mod(L,10)=6$.

\emph{Derived construction (J8-6)}: If $mod(L,10)=6$, then set $\textbf{E}=[0,8,L+1,L+17,3L+11,3L+27,4L+20,4L+28]^T\cdot[0,1,\cdots,L-1]$.

\subsection{Lower Bound on Consecutive Circulant Sizes}
By a similar reasoning as in the proof of Lemma 1 or 2, the following two lemmas can be proved. The detailed proofs are omitted for the sake of brevity.

\emph{Lemma 8}: Let $mod(L,10)\in\{1,3,5\}$. For each circulant size $P\geq(L-1)(4L+12)+1$, the Tanner graph associated with $\textbf{E}$ in Basic Construction (J8-135) has a girth of eight.

\emph{Lemma 9}: Let $mod(L,10)=7$. For each circulant size $P\geq(L-1)(4L+24)+1$, the Tanner graph associated with $\textbf{E}$ in Basic Construction (J8-7) has a girth of eight.

\emph{Remark 3}: By referring to Section III. A, it is readily seen that the values of $\textbf{M}(J-1)\cdot(L-1)+1$ for DC(J8-024), DC(J8-9), DC(J8-8) and DC(J8-6) are $(4L+16)(L-1)+1$, $(4L+20)(L-1)+1$, $(4L+24)(L-1)+1$ and $(4L+28)(L-1)+1$, respectively.

From Lemmas $8\sim9$ and Remark 3, a theorem follows immediately regarding the existence of $(8,L)$-regular QC-LDPC code without 4-cycles and 6-cycles.

\emph{Theorem 3}: There exist $(8,L)$-regular QC-LDPC codes with a girth at least eight for each row weight $L>8$ and each circulant size $P\geq P_{LB}=4(L-1)(L+7)+1$.

\subsection{Circulant Size Smaller Than Lower Bound}
By a similar reasoning as in the proof of Lemma 3, the following four lemmas can be proved. The detailed proofs are omitted for the sake of brevity.

\emph{Lemma 10}: If $mod(L,10)=\{1,3\}$, the Tanner graph associated with \textbf{E} of BC(J8-135) and the circulant size $P=L(3L+8)$ has a girth of eight.

\emph{Lemma 11}: If $mod(L,10)=5$, the Tanner graph associated with \textbf{E} of BC(J8-135) and the circulant size $P=L(3L+10)$ has a girth of eight.

\emph{Lemma 12}: If $mod(L,10)=7$, the Tanner graph associated with \textbf{E} of BC(J8-7) and the circulant size $P=L(3L+16)$ has a girth of eight.

\emph{Lemma 13}: If $mod(L,10)=9$, the Tanner graph associated with \textbf{E} of DC(J8-9) and the circulant size $P=3L^2+16L+12$ has a girth of eight.

\emph{Remark 4}: In Lemma 10, choose the first $L-1$ columns of $\textbf{E}$. Then the obtained matrix corresponds to a $(8,L-1)$-regular QC-LDPC code with a girth of eight for the circulant size of $P=L(3L+8)$. By setting $L=L'+1$ where $mod(L',10)\in\{0,2\}$, it is easy to see that $(8,L')$-regular QC-LDPC code with a girth of eight can be constructed for $P=(L'+1)(3L'+11)$. Similarly, according to Lemma 11, $(8,L')$-regular QC-LDPC code with a girth of eight can be constructed for $P=(L'+1)(3L'+13)$, where $mod(L',10)=4$. Again, due to Lemma 12, $(8,L')$-regular QC-LDPC code with a girth of eight can be generated for $P=(L'+1)(3L'+19)$, where $mod(L',10)=6$. Finally, owing to Lemma 13, $(8,L')$-regular QC-LDPC code with a girth of eight can be obtained for $P=3L^2+22L+31$, where $mod(L',10)=8$.

From Lemmas $10\sim 13$ and Remark 4, a theorem follows immediately regarding the existence of $(8,L)$-regular QC-LDPC codes without 4-cycles and 6-cycles. This theorem provides an upper bound on the minimum circulant size.

\emph{Theorem 4}: For each row weight $L>8$, the minimum circulant size $P_s$ for $(8,L)$-regular QC-LDPC codes with a girth at least eight satisfies $Ps\leq P_{UB}=(3L+4)(L+6)+7$.

According to Theorem 4, if one needs to find out the minimum circulant size for a $(8,L)$-regular QC-LDPC code without 4-cycles and 6-cycles, the search space can be narrowed down to the range between $7(L-1)+1$ and $P_{UB}$.

\section{Comparison with Existing Benchmarks Regarding Circulant Size}

\begin{figure}[t]
\centering
\includegraphics[width=6in]{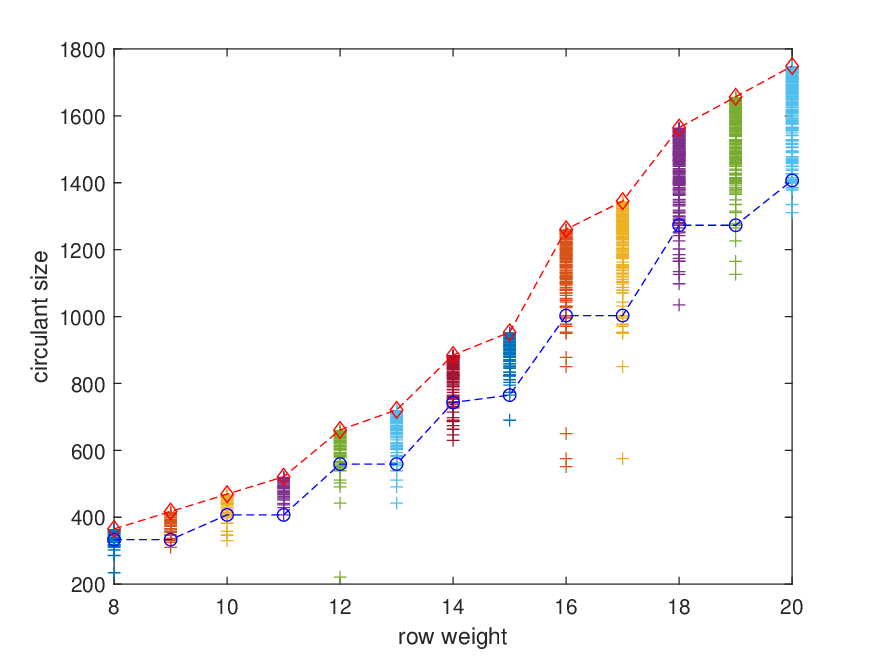}
\caption{Circulant sizes guaranteeing girth-8 $(7,L)$-regular codes: explicitly determined value (dotted blue line) and all empirical values (marked with `+') below the lower bound on consecutive circulant sizes (dotted red line).}
\label{fig1}
\end{figure}

\begin{figure}[t]
\centering
\includegraphics[width=6in]{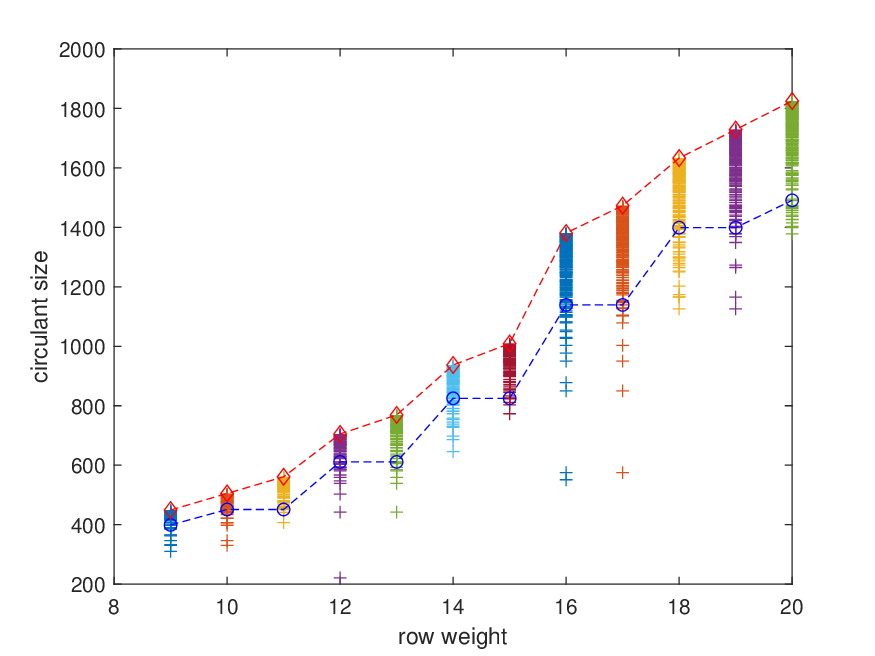}
\caption{Circulant sizes guaranteeing girth-8 $(8,L)$-regular codes: explicitly determined value (dotted blue line) and all empirical values (marked with `+') below the lower bound on consecutive circulant sizes (dotted red line).}
\label{fig2}
\end{figure}

A theorem on the existence of girth-g $(J,L)$-regular QC-LDPC codes is proposed in \cite{JF06}, which states that: when the circulant size $P\geq [(J-1)(L-1)]^{g/2}/(JL-J-L)$, a girth-g $(J,L)$-regular QC-LDPC code definitely exists. Therefore, the bound for $J=7$ and $g=8$ can be readily computed as $6^4\cdot(L-1)^4/(6L-7)\approx 216\cdot(L-1)^3$. Later, this bound is improved to $(L^2+L)(L-1)+1$ \cite{ZZ14} and further to $[L(L-3)+2](L-1)+1$ \cite{MGR22}. The foregoing bounds are all approximately the cube of $L$. Recently, this bound is significantly improved to $(7L-4)(L-1)+1$ \cite{ZLSF24} and further to $5(L+4)(L-1)+1$ \cite{ZFL19}. Thanks to Theorem 1 in this paper, this bound is noticeably improved to $4(L+5)(L-1)+1$, asymptotically about $20\%$ enhancement over the current best result.

Similarly, regarding $J=8$, the bound presented in \cite{JF06} can be readily computed as $7^4\cdot(L-1)^4/(7L-8)\approx 343\cdot(L-1)^3$. Later, this bound is improved to $(L^2+L+1)(L-1)+1$ \cite{ZZ14}. The two bounds are all approximately the cube of $L$. Recently, this bound is significantly improved to $(5L+23)(L-1)+1$ \cite{ZFL19}. According to Theorem 3 in this paper, this bound is noticeably improved to $4(L+7)(L-1)+1$, also asymptotically about $20\%$ enhancement over the current best result.

Furthermore, Theorem 2 in this article constructively proves the existence of girth-8 $(7,L)$-regular QC-LDPC codes, whose circulant sizes are asymptotically $25\%$ lower than the bound stated in Theorem 1. Theorem 4 in this article also constructively proves the existence of girth-8 $(8,L)$-regular QC-LDPC codes, whose circulant sizes are again asymptotically $25\%$ lower than the bound stated in Theorem 3.

Fig.~\ref{fig1} depicts all feasible circulant sizes for $J=7$ provided by the above basic and derived constructions in Section II. A. The red dashed line represents the lower bound $P_{LB}*$ given by Lemmas $1\sim 2$ and Remark 1, and the blue dashed line shows the circulant size $P_s$ determined by the formulae in Lemmas $3\sim 7$ and Remark 2. Note that all values not less than $P_{LB}*$ can be considered as circulant sizes, so they are not marked. From the figure, it can be seen that: (1) the circulant size $P_s$ below $P_{LB}*$ is indeed very small; (2) many circulant sizes below $P_{LB}*$ cannot be uniformly expressed by formulae, but they are even smaller. Fig.~\ref{fig2} illustrates the case of $J=8$, and the properties on circulant sizes are similar to those in Fig.~\ref{fig1}.

\section{Performance Comparison}

\begin{figure}[t]
\centering
\includegraphics[width=6in]{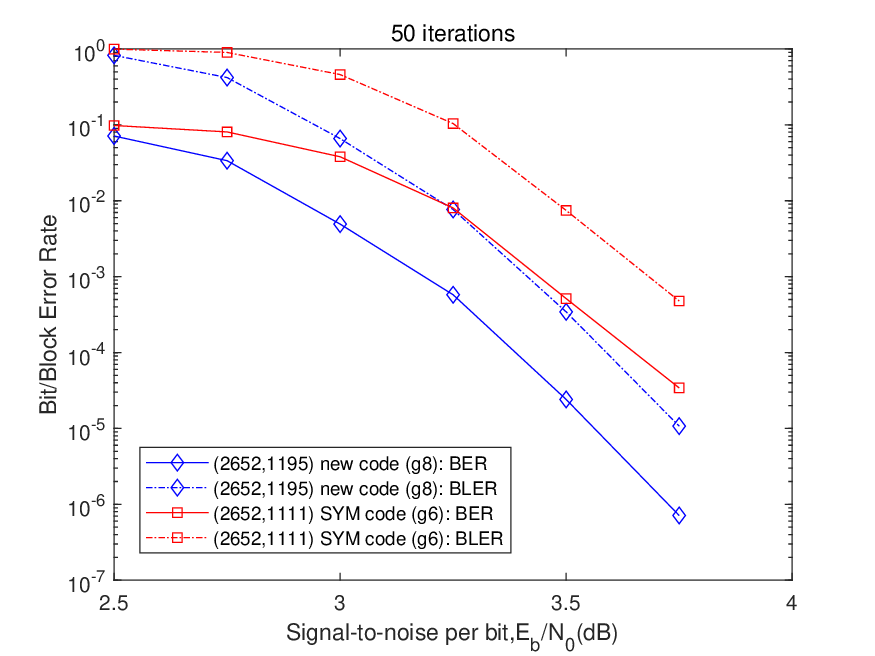}
\caption{Performance comparison of new code (girth 8) and symmetrical code (girth 6) with circulant size of 221.}
\label{fig3}
\end{figure}

\begin{figure}[t]
\centering
\includegraphics[width=6in]{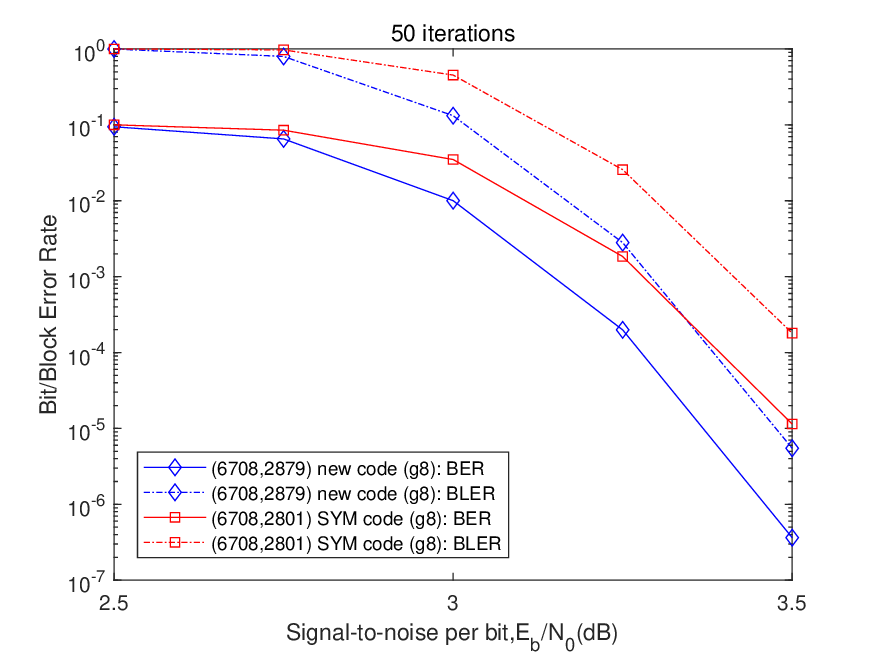}
\caption{Performance comparison of new code (girth 8) and symmetrical code (girth 8) with circulant size of 559.}
\label{fig4}
\end{figure}

This section uses two examples with $J=7$ and $L=12$ to illustrate the performance advantages of codes obtained by the novel algebraic methods compared to codes obtained by a typical search method with symmetrical structure (named SYM code \cite{TBS17}). In the first example, the circulant sizes are both 221, but the new code has a girth of 8 while the SYM method can only find out the code with a girth of 6. In the second example, the circulant sizes are both 559, and they both have a girth of 8. It is observed in Fig.~\ref{fig3} and Fig.~\ref{fig4} that the algebraically constructed new codes significantly outperform the search-based counterparts with the SYM structure.

\end{document}